\documentclass[review]{elsarticle}

\usepackage{lineno,hyperref}

\usepackage{dcolumn}
\newcolumntype{d}[1]{D{.}{.}{#1}}

\usepackage{amssymb}
\usepackage{bbold}
\usepackage{graphicx}
\usepackage{tabularx}
\usepackage[utf8]{inputenc}
\usepackage[english]{babel}
 \usepackage{amsthm}
\usepackage{amsmath,amsfonts}
\usepackage[usenames, dvipsnames]{color}
\usepackage{dsfont}
\usepackage{bbm}
\usepackage{breqn}
\usepackage{amssymb}
\modulolinenumbers[5]

\theoremstyle{definition}
\newtheorem{definition}{Definition}
\newtheorem{theorem}{Theorem}

\newtheorem{lemma}[theorem]{Lemma}

\usepackage{natbib}

\usepackage{xcolor}

\journal{Journal of \LaTeX\ Templates}

\begin{document}

\begin{frontmatter}

\title{A  spatio-temporal multi-scale model for Geyer saturation point process: application to forest fire occurrences
}



\author[mymainaddress]{Morteza Raeisi\corref{mycorrespondingauthor}}
\cortext[mycorrespondingauthor]{Corresponding author}
\ead{morteza.raeisi@univ-avignon.fr}
\author[mymainaddress,mysecondaryaddress]{Florent Bonneu}
\author[mysecondaryaddress]{Edith Gabriel}

\address[mymainaddress]{LMA EA2151, Avignon University, F-84000 Avignon, France}
\address[mysecondaryaddress]{INRAE, BioSP, F-84914 Avignon, France}

\begin{abstract}
Because most natural phenomena  exhibit dependence at multiple scales like locations of earthquakes or forest fire occurrences, spatio-temporal single-scale point process models are unrealistic in many applications.
This  motivates us to construct generalizations of classical Gibbs models. In this paper, we extend the Geyer saturation point process model to the spatio-temporal multi-scale framework. The simulation process is carried out through a birth-death Metropolis-Hastings algorithm.
In a simulation study, we compare two common methods for statistical inference in Gibbs models: the pseudo-likelihood  and logistic likelihood approaches that we tailor to this model. Finally, we illustrate this new model on forest fire occurrences modelling in Southern France.
\end{abstract}


\end{frontmatter}

\section{Introduction}

Nowadays point process models are widely used to highlight trends and interactions in the spatial or spatio-temporal distribution of events.
Most of them are single-structure in the sense that they exhibit either spatial randomness (e.g. modelled by the Poisson process~\citep{kingman1993,kingman2006}) or clustering (mostly modelled by Cox processes~\citep{cox1972}, in particular log-Gaussian Cox processes~\citep{moller1998,brix2001a,brix2001b,diggle2013a}, Poisson Cluster processes~\citep{neyman1958,brix2002,gabriel2014} and Shot-Noise Cox processes~\citep{brix2000,moller2004,moller2010}) or inhibition (modelled by Strauss processes~\citep{strauss1975,cronie2015}, Mat{\'e}rn hard core  processes~\citep{matern1960,gabriel2013} and determinantal point processes~\citep{macchi1975, lavancier2015}).
However, lot of phenomena present interactions at different  scales what motivate statisticians to develop new models, mainly spatial models in ecology~\cite{levin1992,wiegand2007,picard2009}, epidemiology~\cite{iftimi2017} or seismology~\cite{siino2017,siino2018b}, but very few spatio-temporal models in environment~\cite{gabriel2017} or epidemiology~\cite{iftimi2018} as lately reviewed in  \cite{raeisi2019}.
Multi-scale models are mostly based on Gibbs models (see~\cite{dereudre2019} for a recent review on Gibbs models) as they offer a large class of models which allow any of the above mentioned interaction structure. Multi-structure models can then be obtained by hybridization~\cite{baddeley2013}.

Gibbs point processes are studied by their probability density, defined with respect to the unit rate Poisson point process.
Well-known  inhibitive  Gibbs models include the hardcore model (events are forbidden to come too close together) and the Strauss model~\cite{strauss1975} (pairs of close events are not impossible but are unlikely to occur). Generalizing the Strauss process, the Geyer saturation process~\cite{geyer1999} intends to model both inhibition and clustering.
It is able to take into account the clustering nature of a pattern due to interactions between points in absence of covariate information~\cite{anwar2015}.

\cite{baddeley2013} defined a new class of multi-scale Gibbs point processes, so-called hybrid models. The hybridization technique consists in defining the density function of a multi-scale point process model as the product of several densities of Gibbs point processes, $f_l$ for $l=1,\dots,m$, so that
$f=cf_1\times \cdots \times f_m$ 
where $c$ is a normalization constant.
The choice of the normalization constant allows to well define a probability density in the case where the product of densities is integrable.
In particular, \cite{baddeley2013} introduced the spatial multi-scale Geyer saturation point process that has then been applied in epidemiology~\cite{iftimi2017}  and in seismology~\cite{siino2017,siino2018b}. 
\cite{iftimi2018} extended the hybridization approach to the spatio-temporal framework and introduced the spatio-temporal multi-scale area-interaction 
process. New models remain to be developed in the spatio-temporal framework to better describe complex phenomena.

Forest fire occurrences present multi-scale structures which are related to spatial or spatio-temporal inhomogeneities of environmental and climate covariates as well as influence of past events. Their complex interaction structure has been modelled by a spatio-temporal log-Gaussian Cox process in~\cite{opitz2020} and with an inhibitive effect as covariate in~\cite{gabriel2017}.
Gibbs point process models have also been considered in the spatial context for modelling wildfires like the area-interaction point process~\cite{juan2012,serra2013,trilles2013,arago2016,woo2017} or the Geyer point process~\cite{turner2009}. 
In this paper, we aim to introduce the spatio-temporal multi-scale Geyer saturation point process for modelling forest fire occurrences. Our data, available from the Prométhée database\footnote{{\tt https://www.promethee.com/en}}, concern forest fire occurrences in the Bouches-du-Rh{\^o}ne department  (South of France) between 2001 and 2015.

Our model is introduced in Section~\ref{sect.3}. 
In Section \ref{sect.4},  we extend the pseudo-likelihood and logistic likelihood approaches for statistical inference of Gibbs models to the spatio-temporal framework. Then in Section~\ref{sect.5} we implement the model simulation using a birth-death Metropolis-Hastings algorithm and present a simulation study   to compare the performance of the two estimation methods. Finally, in Section~\ref{sect.6}, we apply our model to forest fire occurrences in Southern France.

\section{Spatio-temporal Geyer saturation point process}   \label{sect.3}

A spatio-temporal point process can be viewed as a random locally finite subset of a Borel set $W=S \times T \subset \mathbbm{R}^2\times \mathbbm{R}$. We consider a complete, separable metric space $(W,d(\cdot,\cdot))$  where $d((u,v),(u',v')):=\max\{||u-u'||,|v-v'|  \}$ for $(u,v),(u',v') \in W$. For ${\cal N}$ the state space of points configurations of $W$, $\textbf{x} \in {\cal N}$ denotes a point pattern, i.e. $\textbf{x}=  \{(\xi_1,t_1),...,(\xi_n,t_n)\}$ where $(\xi_i,t_i)$ describes the location and  time, respectively, associated with the $i^{th}$ event. 

The cylindrical neighbourhood $C_r^q(u,v)$ centred at $(u,v)\in S \times T$ is defined as
\begin{equation}
C_r^q(u,v)=\{ (a,b)\in S \times T : ||u-a||\leq r ,|v-b|\leq q \},
\end{equation}
where $r,q>0$ are spatial and temporal radii,  $||\cdot||$ denotes the Euclidean distance in $\mathbbm{R}^2$ and $|\cdot|$ denotes the usual distance in $\mathbbm{R}$. Note that $C_r^q(u,v)$ is a cylinder
with centre $(u,v)$, radius $r$, and height $2q$ that represents a natural neighbourhood for extending spatial Gibbs models to the spatio-temporal context \citep{gonzalez2016}.

The Papangelou conditional intensity \citep{papangelou1974} of a spatio-temporal point process on $W$ with density $f$ is defined by 
\begin{equation}
\lambda((u,v)|\textbf{x}) = \frac {f(\textbf{x} \bigcup  (u,v))}{f(\textbf{x}\backslash (u,v))},
\end {equation}
with $a/0:= 0$ for $a \geq 0$ and $(u,v)\in  W$ (\cite{cronie2015}). Hence, we have $\lambda((u,v)|\textbf{x}) = \frac {f(\textbf{x} \bigcup  (u,v))}{f(\textbf{x})}$ if $(u,v)\notin \textbf{x}$ and  $\lambda((u,v)|\textbf{x}) = \frac {f(\textbf{x})}{f(\textbf{x} \backslash (u,v))}$ if $(u,v)\in \textbf{x}$. 

\cite{gonzalez2016}  introduced a spatio-temporal Strauss process with conditional intensity for $(u,v)\notin \textbf{x}$
\begin{equation}
\lambda ((u,v)|\textbf{x})=\lambda  \gamma ^{n[C_r^q(u,v);\textbf x]},
\label{Eq.6}
\end{equation}
where $n[C_{r}^{q}(u,v);\textbf{x}]=\sum _ {(\xi,t)\in \textbf{x} }  \mathbb{1} \{||u-\xi||\leq r, |v-t|\leq q\}$ is the number of points of $\textbf{x}$ 
lying in $C_r^q(u,v)$. 

 The density function of Strauss model is not integrable for $\gamma > 1$,  it thus does not define a valid probability density and the Strauss process can not be intended for clustering structures. To avoid this issue, \citep{geyer1999} consider an upper bound (saturation parameter) for the number of neighboring points that interact and define the (spatial) Geyer saturation point process.

\theoremstyle{definition}
\begin{definition}
We define the \textit{spatio-temporal Geyer saturation point process} as the point process  with density
\begin{equation}
f(\textbf{x})=c \prod_{(\xi,t)\in \textbf{x}} \lambda(\xi,t)  \gamma^ {\min\{s,n(C_{r}^{q}(\xi,t);\textbf{x})\}},
\label{Eq.07}
\end{equation}
with respect to a unit rate Poisson process on $W$, where $c > 0$ is a normalizing constant, $\lambda$ is a non-negative, measurable and bounded function, 
$\gamma>0$ is
the interaction parameter, $s$ is the saturation parameter, and $n(C_{r}^{q}(\xi,t);\textbf{x})=\sum _ {(u,v)\in \textbf{x} \backslash (\xi,t)}  \mathbbm{1} (||u-\xi||\leq r, |v-t|\leq q)$ is the number of points of $\textbf{x}$ lying in $C_{r}^{q}(\xi,t)$ and different from $(\xi,t)$.
\end{definition}

The function $\lambda$ describes some spatio-temporal trend in point pattern that can be estimated using covariates. 
The scalars $ \gamma, r,q$ and $s$  are the parameters of the model. The saturation parameter $s$ is an upper bound of the number of points 
 in the cylinder $C_r^q$. By using hybridization approach \citep{baddeley2013,iftimi2018}, we  define a  multi-scale  version of (\ref{Eq.07}).
\theoremstyle{definition}
\begin{definition} \label{def2}
We define the \textit{spatio-temporal multi-scale  Geyer saturation point process} as the point process with density
\begin{equation}
f(\textbf{x})=c \prod_{(\xi,t)\in \textbf{x}} \lambda(\xi,t) \prod_{j=1}^{m}\gamma_j ^ {\min\{s_j,n(C_{r_j}^{q_j}(\xi,t);\textbf{x})\}},
\label{Eq.7}
\end{equation}
with respect to a unit rate Poisson process on $W$, where  $\gamma_j> 0$, $j=1,\dots,m$, are the interaction parameters, and $r_1<\cdots<r_m$, $q_1<\cdots<q_m$ are spatial and temporal interaction ranges. 
\end{definition}
For any $j \in \lbrace 1, . . . ,m \rbrace$, the interaction parameters $0 <
\gamma_j < 1$ traduce inhibition, while $\gamma_j > 1$ traduce clustering between  points at some spatio-temporal scales. 
When  $s_j = 0$ or $\gamma_j=1$  for all
$j \in \lbrace 1, . . . ,m \rbrace$, the density  (\ref{Eq.7}) 
reduces to the one of an inhomogeneous Poisson process. 
Equation~(\ref{Eq.7}) indicates that the structure of the process changes with the spatial and temporal distances $r_j,q_j$. 
Covariates can be added to the model by assuming that the spatio-temporal trend $\lambda$ is function  of a covariate vector $Z(\xi, t)$, i.e. $\lambda(\xi, t) = \Psi(Z(\xi, t))$.

\theoremstyle{lemma}
\begin{lemma} \label{lem1}
The spatio-temporal multi-scale Geyer point process  is a Markov point process in the sense of Ripley-Kelly~\citep{ripley1977} and  its density (\ref{Eq.7}) is measurable and integrable for all $\gamma_j,$ $j = 1, \dots , m$ with $m \in \mathbbm{N}$.
\end{lemma}
\begin{proof}
A Geyer model is hereditary, locally  and Ruelle stable and hence integrable~\citep{geyer1999}. ~\citep{baddeley2013} showed these properties for hybrids. 
As in~\cite{iftimi2018}, we can show that the spatio-temporal Geyer saturation point process (\ref{Eq.07}) is a Markov point process in Ripley-Kelly’s sense at interaction range $2 \max\{r, q\}$ and  that the spatio-temporal  multi-scale  Geyer saturation process (\ref{Eq.7}) is also a Markov point process in Ripley-Kelly sense at interaction range $\max_{1\leq j \leq m}\{2 \max\{r_j, q_j\}\}=2 \max\{r_m, q_m\}$ (\cite{baddeley2013}). 
\end{proof}
For any $(u,v) \in W $, the Papangelou conditional intensity function of the spatio-temporal multi-scale  Geyer saturation process is 
\begin{align}
\begin{split}
\lambda((u,v)|\textbf{x})&= \lambda (u,v) \prod_{j=1}^{m} \gamma_j ^{\min \{s_j,n(C_{r_j}^{q_j}(u,v);\textbf{x})\}} \\
&\times \prod_{(\xi,t)\in \textbf{x} \setminus (u,v)} \gamma_j ^{\min \{s_j,n(C_{r_j}^{q_j}(\xi,t));\textbf{x} \cup (u,v))\}-\min \{s_j,n(C_{r_j}^{q_j}(\xi,t);\textbf{x}\setminus (u,v))\}},
\end{split}
\label{Eq.100}
\end{align}
The  Markovian property (Lemma \ref{lem1}) ensures that 
this conditional intensity only depends on $(u,v)$ and its neighbors in $\textbf{x}$. Hence, we can design simulation algorithms for generating  realizations of the model, see Section~\ref{sect.5}.

\section{Inference}   \label{sect.4}

Geyer saturation point process model  (\ref{Eq.07}) involves two types of parameters: regular parameters
and irregular parameters. A parameter is called {\it regular} if the log likelihood is a linear function of that parameter, {\it irregular} otherwise. Regular parameters like  trend $\lambda$ and interaction $\gamma$
can be estimated using the pseudo-likelihood method \citep{baddeley2000} or the logistic likelihood method \citep{baddeley2014} rather than the maximum likelihood method \citep{ogata1981}. Indeed, they are based on the conditional intensity (and hence are free from the normalization constant $c$) and  easily computable for most Gibbs models. Here we consider these two methods to estimate regular parameters and we compare their performance in the next subsection.

Irregular parameters, like saturation threshold $s$ and distances $r$ and $q$, are difficult to estimate using the maximum likelihood method because the likelihood function is not differentiable with respect to them. 
These parameters can be estimated using the profile pseudo-likelihood approach~\citep{baddeley2000} or predetermined by the user using some summary statistics, like the pair correlation and the auto-correlation functions~\citep{iftimi2018}, in order to determine the interaction ranges. ~\cite{baddeley2006} presented the methods that are used for irregular parameter estimation in the spatial framework.

In this paper, we combine the advantages of the two previous methodologies. By computing some statistics summarizing the range of interactions in space and time, we consider a set of feasible irregular parameter values and we choose the combination of them providing the best Akaike's Information Criterion (AIC) for the fitted model. 

\subsection{Pseudo-likelihood approach}

Let $\boldsymbol \theta$ be the vector of regular parameters that we aim to estimate.
~\cite{besag1977} defined the pseudo-likelihood for spatial point processes in order
to avoid computational problems with point process likelihoods. One can easily extend it for a spatio-temporal point process with conditional intensity 
$\lambda_{\boldsymbol \theta}((u,v)| \textbf{x})$  over $W$ as follows
\begin{equation}
PL(\textbf{x};\boldsymbol \theta)=\exp \left(-\int_{S} \int_{T} \lambda_{\boldsymbol\theta}((u,v)|\textbf{x})dvdu \right) \prod_{(\xi,t)\in \textbf{x}}\lambda_{\boldsymbol\theta}((\xi,t)|\textbf{x}).
\label{Eq.14}
\end{equation}
The pseudo score is defined by
\begin{equation}
U(\textbf{x};\boldsymbol \theta)=\frac {\partial       }{\partial      \boldsymbol   \theta}\log PL(\textbf{x};\boldsymbol\theta),
\end{equation}
that is an unbiased estimating function. The maximum pseudo-likelihood normal equations are then given by
\begin{equation}
\frac {\partial       }{\partial      \boldsymbol   \theta}\log PL(\textbf{x};\boldsymbol\theta)=0,
\label{Eq.15}
\end{equation}
where
\begin{equation}
\log PL(\textbf{x};\boldsymbol\theta)= \sum_{(\xi,t)\in \textbf{x}}\log \lambda_{\boldsymbol\theta}((\xi,t)|\textbf{x}) -\int_{S} \int_{T} \lambda_{\boldsymbol\theta}((u,v)|\textbf{x})dvdu,
\label{Eq.16}
\end{equation}
and $\lambda_{\boldsymbol\theta}(\cdot|\textbf{x})$ is defined by (\ref{Eq.100})  for hybrid Geyer model (\ref{Eq.7}). 

For sake of clarity, we now assume that 
$\boldsymbol\theta = [\log \gamma_1,\dots, \log \gamma_m]^{\top}$
 the logarithm of interaction parameters in model (\ref{Eq.7}). To estimate  $ \boldsymbol\theta$, we  use the pseudo-likelihood approach.
Equation~(\ref{Eq.100}) can be rewritten as 
 $\lambda_{\boldsymbol\theta}((u,v)|\textbf{x})=\lambda (u,v) \prod_{j=1}^{m} \exp(\theta_j S_j ((u,v),\textbf{x}))$ where
\begin{align}
\begin{split}
 S_j ((u,v),\textbf{x})&=\min \{s_j,n(C_{r_j}^{q_j}(u,v);\textbf{x})\} \\
&\hspace{.5cm}+ \sum_{(\xi,t) \in \textbf{x} \backslash (u,v) }[\min \{s_j,n(C_{r_j}^{q_j}(\xi,t);\textbf{x} \cup (u,v))\}\\
&\hspace{2.6cm}-\min \{s_j,n(C_{r_j}^{q_j}(\xi,t);\textbf{x} \backslash (u,v) )\}],
\end{split}
\label{Eq.17}
\end{align}
is a sufficient statistics.
Then, for $\boldsymbol S((u,v),\textbf{x})=[S_1((u,v),\textbf{x}), \dots,S_m((u,v),\textbf{x})]^{\top}$
\begin{equation}
\log \lambda_{\boldsymbol\theta}((u,v)|\textbf{x}) =\log \lambda (u,v) +\boldsymbol \theta^{\top} \boldsymbol S((u,v),\textbf{x})
\label{Eq.18}
\end{equation}
is a linear model in $\boldsymbol\theta$ with offset $\log \lambda(u,v)$.
Thus, equation~(\ref{Eq.15}) gives us the pseudo-likelihood equations
\begin{equation}
\dfrac {\partial       }{\partial     \boldsymbol    \gamma} \left\lbrack \sum_{(\xi,t)\in \textbf{x}}  [\log \lambda(\xi,t) +\sum_{j=1}^{m} \log (\gamma_j)S_j((\xi,t),\textbf{x}) ]
-\int_{S} \int_{T} \lambda (u,v) \prod _ {j=1}^{m} \gamma_j ^ {S_j ((u,v), \textbf{x})}dvdu \right\rbrack=0,
\label{Eq.19}
\end{equation}
For each parameter $\gamma_i$, $i = 1,\dots ,m$, the equations (\ref{Eq.19}) can be rewritten
\begin{equation}
\sum_{(\xi,t)\in \textbf{x}} \frac {S_i((\xi,t),\textbf{x})}{\gamma_i}=\int_{S} \int_{T} \lambda (u,v) \frac {S_i((u,v),\textbf{x})}{\gamma_i} \prod _ {j=1}^{m} \gamma_j ^ {S_j ((u,v),\textbf{x})}dvdu,
\label{Eq.21}
\end{equation}
The major difficulty is to estimate the integrals on the right hand side of equations
(\ref{Eq.21}).  The pseudo-likelihood cannot be computed exactly but
must be approximated numerically. 

For a point process model, the approximation of likelihood is converted into a regression model. In the following, we refer to generalized log-linear Poisson regression approach as approximation of integrals in (\ref{Eq.21}). In the next section, we also investigate an alternative, the logistic regression.

~\cite{berman1992} developed a numerical quadrature method to approximate
maximum likelihood estimation for an inhomogeneous Poisson point process. Berman-Turner method
has then been extended to Gibbs point processes by~\cite{baddeley2000},  approximating the integral in (\ref{Eq.16}) by a Riemann sum
\begin{equation}
\int_{S} \int_{T} \lambda_{\boldsymbol \theta}((u,v)|\textbf{x})dvdu\approx \sum_{k=1}^{n+p} w_k \lambda_{\boldsymbol \theta}((\xi_k,t_k)|\textbf{x}),
\label{Eq.22}
\end{equation}
where $(\xi_k,t_k)$ are points in $\{(\xi_1,t_1),...,(\xi_n,t_n),(\xi_{n+1},t_{n+1}),...,(\xi_{n+p},t_{n+p})  \}\in W$ consisting of the $n$ events of $\mathbf{x}$ and $p$ dummy points, and $w_k$ are quadrature weights such that $\sum_{k=1}^{n+p}w_k=\ell(S\times T)$ where  $\ell$ is Lebesgue measure.
This yields an approximation 
for the log pseudo-likelihood of the form
\begin{equation}
\log PL(\textbf{x};\boldsymbol\theta) \approx \sum_{(\xi,t)\in \textbf{x}}\log \lambda_{\boldsymbol\theta}((\xi,t)|\textbf{x}) - \sum_{k=1}^{n+p} w_k \lambda_{\boldsymbol\theta}((\xi_k,t_k)|\textbf{x}).
\label{Eq.23}
\end{equation}
Note that if the set of points $\{(\xi_k,t_k), k = 1, \dots, n+p\}$ includes all the points of $  \textbf{x}=\{(\xi_1,t_1),...,(\xi_n,t_n)\}$, we
can rewrite (\ref{Eq.23}) as
\begin{equation}
\log PL(\textbf{x};\boldsymbol\theta) \approx  \sum_{k=1}^{n+p} w_k \left(y_k \log \lambda_{\boldsymbol\theta}((\xi_k,t_k)| \textbf{x} )-\lambda_{\boldsymbol\theta}((\xi_k,t_k)| \textbf{x} ) \right),
\label{Eq.24}
\end{equation}
where 
\begin{equation}
  y_k=
  \begin{cases}
                                   1/w_k, & \text{if $(\xi_k,t_k)\in \textbf{x} $ is an event,} \\
                                   0, & \text{if $(\xi_k,t_k)\notin \textbf{x} $ is a dummy point.} 
  \end{cases}
\label{Eq.25}
\end{equation}

The right hand side of (\ref{Eq.24}), for fixed $\textbf{x}$, is formally equivalent to the log-likelihood of independent Poisson variables $Y_k \sim Poisson(\lambda_{\boldsymbol\theta}((\xi_k,t_k)| \textbf{x} ))$ taken with weights $w_k$. 
Therefore, by using the \texttt{glm} function in \texttt{R} (\cite{r2016}), we can perform the maximum likelihood-based parameter estimation of this Poisson generalized linear model and obtain the maximum value for~(\ref{Eq.24}).

Note that in hybrid Geyer model (\ref{Eq.7}), we consider $\lambda(\xi,t)=\lambda_{\beta}(\xi,t)=\beta \mu(\xi,t)$ where $\mu(\xi,t)$ is known or estimated beforehand and $\beta$ is a parameter to estimate. In summary, the method is as follows.

\vspace{.2cm}
\textbf{\textit{Algorithm 1}}
\begin{itemize}

\item Generate a set of $p$ uniform dummy points in $W$ and merge them with all the data points
in $\textbf{x}$ to construct the set of quadrature points $(\xi_k,t_k) \in W$ with $k=1,\dots,n+p$.

\item Compute the quadrature weights $w_k$ and the indicators $y_k$ defined in (\ref{Eq.25}), 

\item  Compute the sufficient statistics $\boldsymbol S((\xi_k,t_k),\textbf{x})$ at each quadrature point,

\item Fit a  log-linear Poisson regression with explanatory variables $\boldsymbol S((\xi_k, t_k),\textbf{x})$, and offset $\log  \lambda(\xi_k,t_k)$ 
 on the  responses $y_k$ with weights $w_k$ to obtain estimates $\hat{\boldsymbol \theta}$ for the $\boldsymbol S$-vector and intercept $\hat{\theta}_0$,

\item Return the maximum pseudo-likelihood-based parameter estimates $\hat{\gamma_j}=\exp (\hat{\theta_j})$ for $j=1,\dots,m$ and $\hat{\beta}=\exp(\hat{\theta}_0)$.
\end{itemize}
We define the quadrature scheme by defining a spatio-temporal partition of $W$ into cubes $C_k$ of equal volumes $\nu$ and by using the counting weights proposed in~\cite{baddeley2000}. We then assign to each dummy or data point $(\xi_k,t_k)$ a weight $w_k = \nu/n_k$
where  $n_k$ is the number of dummy and data points that lie in the 
same cube as $(\xi_k,t_k)$. An alternative way to define the quadrature
scheme for \textit{Algorithm 1} is based on Dirichlet tessellation~\citep{baddeley2000} and the weight of each point is equal to the volume of the corresponding Dirichlet 3D cell. In this paper, we consider cubes because it is less time consuming and provides similar results (see~\cite{opitz2009} for quadrature schemes comparison of 3D Gibbs point processes).

\subsection{Logistic likelihood approach}

The logistic likelihood method \citep{baddeley2014} is an alternative for estimating the regular parameters of Gibbs models that is closely related to the pseudo-likelihood method. The Berman-Turner approximation often requires a quite large number of dummy points. Hence, fitting such  GLM can be computationally intensive, especially when dealing with a large dataset. 
~\cite{baddeley2014} formulated the pseudo-likelihood estimation equation as a logistic regression using auxiliary dummy point configurations and  
proposed  a computational technique for fitting
Gibbs point process models to spatial point patterns. 
\cite{iftimi2018} extended the logistic likelihood approach for spatio-temporal point processes and we tailored it to our model.

Let $\textbf x$ be a realization of a spatio-temporal point process defined on $W$ having a density $f_{\boldsymbol \theta}$ with respect to the unit rate Poisson process and with conditional intensity function $\lambda_{\boldsymbol \theta}(\cdot|\textbf x )$.
We consider an independent Poisson process for dummy points, with intensity function $\rho$, and we denote by $\boldsymbol d$ a set of dummy points.

By defining $Y(\xi,t)=\mathbbm{1}_{\{(\xi,t)\in \textbf x\}}$ for $(\xi,t)\in \textbf x\cup \boldsymbol d$, we obtain independent Bernoulli variables taking one for data points and zero for dummy points. We have
\begin{eqnarray}
Pr(Y(\xi,t)=1)& = &\frac {\lambda_{\boldsymbol \theta}((\xi,t)|\textbf x \backslash (\xi,t))}{\lambda_{\boldsymbol \theta}((\xi,t)|\textbf x \backslash (\xi,t))+\rho (\xi,t)},
\end{eqnarray}

By considering the log linearity assumption for the conditional intensity $\lambda_{\boldsymbol \theta}(\cdot|\textbf  x)$ in~\eqref{Eq.18}, the logit of $Pr(Y(\xi,t)=1)$ is
\begin{equation}
\log \frac {\lambda_{\boldsymbol \theta}((\xi,t)|\textbf x \backslash (\xi,t))}{\rho (\xi,t)}=\log \frac {\lambda(\xi,t)}{\rho(\xi,t)} + \sum_{j=1}^{m} \theta_j S_j((\xi,t),\textbf x \backslash (\xi,t)),
\label{Eq.2512112}
\end{equation}
which is a linear model in $\boldsymbol \theta$ with offset $\log \frac {\lambda(\xi,t)}{\rho(\xi,t)}$.

Since $\lambda_{\boldsymbol \theta}((\xi,t)|\textbf x )=\lambda_{\boldsymbol \theta}((\xi,t)|\textbf x \backslash (\xi,t))$ for $(\xi,t)\in \boldsymbol d$, the log logistic likelihood is defined by
\begin{align}
\begin{split}
\log LL(\textbf x,\boldsymbol d;\boldsymbol \theta)&=\sum_{(\xi,t)\in \textbf x \cup \boldsymbol d} Y((\xi,t))\log \frac {\lambda_{\boldsymbol \theta}((\xi,t)|\textbf x \backslash (\xi,t))}{\lambda_{\boldsymbol \theta}((\xi,t)|\textbf x \backslash (\xi,t))+\rho (\xi,t)}\\
&+\sum_{(\xi,t)\in \textbf x \cup\boldsymbol d} [1-Y((\xi,t))] \log \frac {\rho(\xi,t)}{\lambda_{\boldsymbol \theta}((\xi,t)|\textbf x )+\rho (\xi,t)}\\
&=\sum_{(\xi,t)\in \textbf x} \log \frac {\lambda_{\boldsymbol \theta}((\xi,t)|\textbf x \backslash (\xi,t))}{\lambda_{\boldsymbol \theta}((\xi,t)|\textbf x \backslash (\xi,t))+\rho (\xi,t)}\\
&+\sum_{(\xi,t)\in \boldsymbol d} \log \frac {\rho(\xi,t)}{\lambda_{\boldsymbol \theta}((\xi,t)|\textbf x )+\rho (\xi,t)}.
\end{split}
\end{align}
The maximum of the log-logistic likelihood exists and under regularity condition is unique~\citep{baddeley2019}. Hence, estimation can be implemented in \texttt{R} by using the \texttt{glm} function.

As in \textit{Algorithm 1}, we consider $\lambda(\xi,t)=\lambda_{\beta}(\xi,t)=\beta \mu (\xi,t)$ and we estimate the regular parameters form the following algorithm.

\vspace{.2cm}
\textbf{\textit{Algorithm 2}}
\begin{itemize}

\item Generate dummy points ${\boldsymbol d}$ from a Poisson process with intensity
function $\rho$ and merge them with all the data points in $\textbf x$ to construct the set of quadrature
points $(\xi_k, t_k) \in W$,

\item Obtain the response variables $y_k$ (1 for data points, 0 for dummy points),

\item  Compute the sufficient statistics $\boldsymbol S((\xi_k,t_k),\textbf{x}\backslash (\xi_k, t_k))$ at each quadrature point,

\item Fit a logistic regression model with explanatory variables $\boldsymbol S((\xi_k, t_k),\textbf{x}\backslash (\xi_k,\\t_k))$, 
and offset $\log \left(\mu(\xi_k, t_k)/\rho(\xi_k, t_k) \right)$ on the  responses $y_k$ to obtain estimates $\hat{\boldsymbol \theta}$ for the $\boldsymbol S$-vector and intercept $\hat{\theta}_0$,

\item Return the parameter estimator $\hat{\boldsymbol \gamma}=\exp (\hat{\boldsymbol \theta})$ and $\hat{\beta}=\exp(\hat{\theta}_0)$ and in the case where $\mu(\xi_k, t_k)/\rho(\xi_k, t_k)$ is a constant $c$ we have $\hat{\beta}=c^{-1} \exp(\hat{\theta}_0)$.

\end{itemize}

\section{Simulation}  \label{sect.5}

The simulation algorithms of Gibbs point process models require only computation of the Papangelou conditional intensity which avoids to consider the difficult estimation of the unknown normalizing constant in the density function. 
Gibbs point process models can be simulated by using Markov chain Monte Carlo (MCMC) algorithms like the birth-death Metropolis-Hastings algorithm~\citep{moller2004} that belongs to the large class of Metropolis-Hastings algorithms~\citep{geyer1994}. 
In this section, we first present the birth-death Metropolis-Hastings algorithm and secondly we investigate the goodness of parameter estimation of the two approaches introduced before.

\subsection{Birth-death Metropolis-Hastings algorithm}  

For $\bf x$ a   spatio-temporal point pattern in $W$, we can propose either a birth  with probability  $q(\textbf{x})$ or a death with probability $1-q(\textbf{x})$. For a birth, a new point $(u,v) \in W$ is
sampled from a probability density $b(\textbf{x}, \cdot)$ and the new point configuration $\textbf{x} \cup (u,v)$ is accepted with
probability $A(\textbf{x}, \textbf{x} \cup (u,v))$, otherwise the state remains unchanged.
For a death, the point $(\xi,t) \in \textbf{x}$
chosen to be removed is selected according to a discrete probability distribution $d(\textbf{x}, .)$ on $\textbf{x}$, and
the proposal $\textbf{x} \setminus (\xi,t)$ is accepted with probability $A(\textbf{x}, \textbf{x} \setminus(\xi,t))$, otherwise the state remains unchanged. For simplicity, we consider $q(\textbf{x})=\frac {1}{2}$, $b(\textbf{x},\cdot)=1/\ell  (W)$ and $d(\textbf{x},\cdot)=1/n(\textbf{x})$. 
By setting $A(\textbf{x}, \textbf{x}\cup (u,v)) = \min\{1, r((u,v);\textbf{x})\}$, and $A(\textbf{x}, \textbf{x} \setminus(\xi,t)) = \min \{1, 1/r((\xi,t);\textbf{x} \setminus (\xi,t))\}$ where $r((u,v);\textbf{x})=\frac{\ell  (W)}{n(\textbf{x})+1}\times \lambda((u,v)|\textbf{x})$ is the Hastings ratio~\citep{iftimi2018}, we obtain the following birth-death Metropolis-Hastings algorithm.

\vspace{.2cm}
\textbf{\textit{Algorithm 3}}
\vspace{.2cm}
For $n=0,1,...,$ given $X_n=\textbf{x}$ (e.g. a Poisson process for $n=0$), generate $X_{n+1}$:
\begin{itemize}

\item Generate two uniform numbers $y_1,y_2$ in $[0,1]$, 

\item If $y_1 \leq  \frac {1}{2} $ then

\begin{itemize}

\item A new point $(u,v)$ is uniformly sampled from a probability density $1/\ell  (W)$,

\item Compute  $r((u,v);\textbf{x})=\frac {\ell  (W)}{n(\textbf{x})+1}\lambda((u,v)|\textbf{x}), (u,v)\notin \textbf{x} $. 

If $y_2<r((u,v);\textbf{x})$ then $X_{n+1}=\textbf{x}  \cup (u,v)$ else $X_{n+1}=\textbf{x}$
\end{itemize}
\item If $y_1 > \frac {1}{2} $ then
\begin{itemize}
\item Uniformly select a point $(\xi,t)$ in $\textbf{x}$ according to a discrete probability density $1/n(\textbf{x})$,
\item Compute $r((\xi,t);\textbf{x}\backslash (\xi,t))=\frac {\ell  (W)}{n(\textbf{x})}\lambda((\xi,t)|\textbf{x} \setminus (\xi,t)) ), (\xi,t)\in \textbf{x}$.

If $y_2<1/r((\xi,t);\textbf{x}\backslash (\xi,t))$ then $X_{n+1}=\textbf{x}  \backslash (\xi,t)$ else $X_{n+1}=\textbf{x}$.
\item Note that if $\textbf{x} = \emptyset $ then $X_{n+1}=\textbf{x}$.
\end{itemize}
\end{itemize}

This simulation process is repeated a large number of time in order to ensure the convergence of the algorithm to the expected distribution. This number of iterations is unknown a priori and must be determined by the user from practical knowledge and/or diagnostic tools. We choose $20,000$ iteration steps in simulation study (70,000 iteration steps in the application study). To investigate the convergence of the algorithm, we use a ``trace plot'' which shows the evolution of the number of points at each iteration of \textit{Algorithm 3}. Thus, we check that the number of points in the simulated point pattern is stabilized (see \cite{moller2004,illian2008} for more details).

\subsection{Simulation  study}  

\begin{figure}
\begin{center}
\includegraphics[width=.3\textwidth]{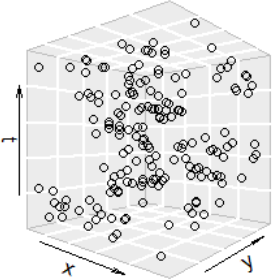}
\includegraphics[width=.3\textwidth]{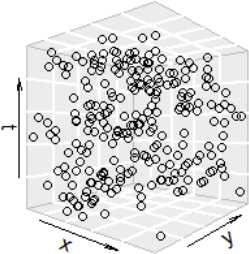}
\includegraphics[width=.3\textwidth]{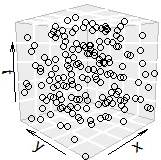}
\caption{\footnotesize{Realizations of \textit{Model 1} (left); \textit{Model 2} (middle); \textit{Model 3} (right).}}
\label{pic1}
\end{center}
\end{figure}

We compare the performance of the pseudo-likelihood and logistic likelihood approaches on the spatio-temporal multi-scale   Geyer point process.
We generate 100 simulated realizations in the unit cube
from three models.
The first one exhibits strong clustering (\textit{Model 1}), the  second one  exhibits small scale inhibition and large scale clustering (\textit{Model 2}) and the  third one  exhibits  inhibition (\textit{Model 3}). 
Model parameters are reported in Table \ref{Tab1.1}. We consider a burn-in period of 20,000 steps in \textit{Algorithm 3}. 
\begin{table}
\centering
\caption{Parameters of the three multi-scale Geyer point process models used in simulation study.}
 \label{Tab1.1}
\scriptsize
\begin{tabular}{ccccccc}
\hline
& \multicolumn{6}{c}{Values of parameter}\\
\cline{2-7}
          &   \multicolumn{2}{c}{Regular parameters}  & & \multicolumn{3}{c}{Irregular parameters}  \\
\cline{2-3}\cline{5-7}
Model & $\lambda$ & $\gamma$  & & $r$ & $q$ & $s$ \\
\hline
\textit{Model 1}   & 70 & (1.5,1.5)   & & (0.05,0.1) &  (0.05,0.1)  &  (2,2)  \\
\textit{Model 2}   & 100 &  (0.5,1.5)   & & (0.05,0.1) &  (0.05,0.1)  &  (1,3) \\
\textit{Model 3}   & 200 & (0.8,0.8)  & & (0.05,0.1) &  (0.05,0.1)  &  (1,1) \\
\hline
\end{tabular}
\end{table}
Figure \ref{pic1} shows one realization of each model.

\begin{figure}
\centering
\includegraphics[width=0.9\textwidth]{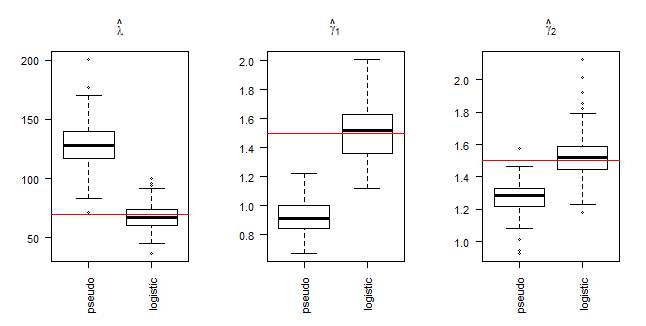}
\includegraphics[width=0.9\textwidth]{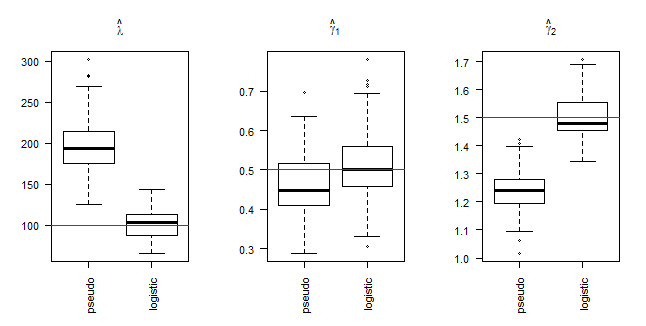}
\includegraphics[width=0.9\textwidth]{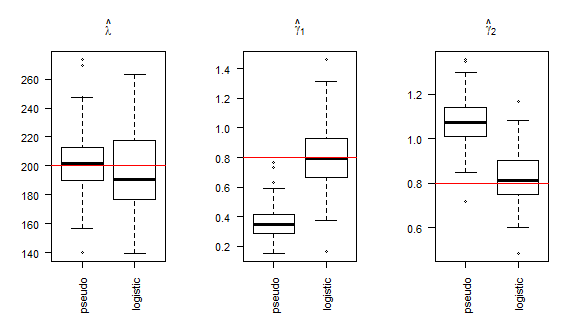}
\caption{Boxplots of regular parameters estimated from the pseudo-likelihood and logistic likelihood  approaches for \textit{Model 1} (first row), \textit{Model 2} (second row) and \textit{Model 3} (third row). True values are represented by horizontal red lines.}
\label{pic2}
\end{figure}

According to~\cite{baddeley2014}, we generate a spatio-temporal Poisson process with intensity $\rho=4n(\textbf {x})$ (resp. $4n(\textbf {x})/ \ell  (W)$) as dummy points in \textit{Algorithm 1} (resp. \textit{Algorithm 2}). 
For each model, we compute the root mean square error (RMSE) of each set of estimated parameters (Table~\ref{Tab1}) and plot the related boxplots (Figure~\ref{pic2}). In Table~\ref{Tab1} the lowest RMSE value is in bold and in Figure~\ref{pic2} the true values are represented by horizontal red lines. Both RMSE and boxplots show that the logistic likelihood approach performs better than the pseudo-likelihood approach for any model. 
\begin{table}
\centering
\caption{RMSE of parameter
estimates from 100 simulated realizations of the multi-scale Geyer point process model.}
 \label{Tab1}
\scriptsize
\begin{tabular}{cccccccccc}
\hline
          &   \multicolumn{3}{c}{\textit{Model 1}}  & \multicolumn{3}{c}{\textit{Model 2}} &
          \multicolumn{3}{c}{\textit{Model 3}} \\
\cline{2-4}\cline{5-7} \cline{8-10}
 Method & $\hat \lambda$ & $\hat  \gamma_1$ & $\hat  \gamma_2$    & $\hat \lambda$ & $\hat  \gamma_1$ & $\hat  \gamma_2$    &  $\hat \lambda$ & $\hat  \gamma_1$ & $\hat  \gamma_2$   \\
\hline
pseudo   & 62.09 & 0.59 & 0.25  &  103.74 &0.09 &  0.27 & \textbf{22.13} & 0.45 &0.29  \\
logistic  & \textbf{12.07} &\textbf{0.18}  & \textbf{0.16} &  \textbf{17.30} & \textbf{0.08} & \textbf{0.08}  &  27.48 &\textbf{0.20}  & \textbf{0.12} \\
\hline
\end{tabular}
\end{table}

Note that in the spatial framework, 
\cite{baddeley2014} showed that for large datasets the 
logistic likelihood method is preferable
than the pseudo-likelihood method
as it requires a smaller number of dummy points and performs quickly and efficiently. ~\cite{daniel2018} and \cite{choiruddin2018}  investigated a
similar comparison when these methods are regularized (i.e. using an approach with a simultaneous parameter estimation and variable selection  by maximizing a penalized likelihood functions). ~\cite{iftimi2018} found the advantage of the logistic likelihood approach for the  spatio-temporal multi-scale area-interaction point process model. 
We here confirm this advantage for the spatio-temporal multi-scale Geyer point process model.

\section {Application to forest fire occurrences}  \label{sect.6}

Economic and ecological disasters caused by wildfires in the world have led the scientific community to develop many novel statistical analysis and modelling wildfire occurrences to better understand their behaviors. 
In this section, we focus on the modelling of forest fire occurrences in the Bouches-du-Rhône county (Southern France) between 2001 and 2015.

Several statistical studies have shown the influence of environmental and meteorological factors on forest fire occurrences. In the French Mediterranean basin, \cite{opitz2020} fit a spatio-temporal log-Gaussian Cox process model for forest fire occurrences with a log-linear intensity depending on spatio-temporal land use and weather covariates.
\cite{ganteaume2013} investigated the impact of the different covariates on the number of fires using multivariate analysis and \cite{gabriel2017} explored the influence of land cover covariates, temperature and precipitation on the probability of event occurrence.
In addition to the spatio-temporal clustering of events induced by some covariates, \cite{gabriel2017} detected spatio-temporal interaction structures at different scales and notably an inhibitive effect that arises locally in time and space after wildfires
as we expect lesser occurrences at these locations during a vegetation regeneration period.

We propose to fit the spatio-temporal hybrid Geyer point process model (\ref{Eq.7}) on wildfire occurrences to take into account both the inhomogeneities induced by covariates and the multi-scale structure of interactions.

\subsection{Data}

\begin{figure}[h]
\begin{center}
\includegraphics[width=6cm,height=5.5cm]{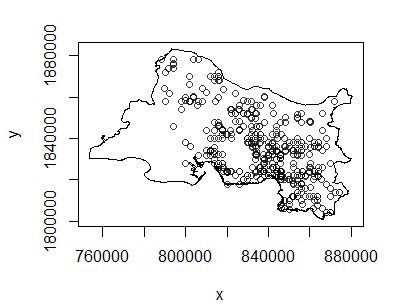}
\includegraphics[width=5.15cm,height=4.85cm]{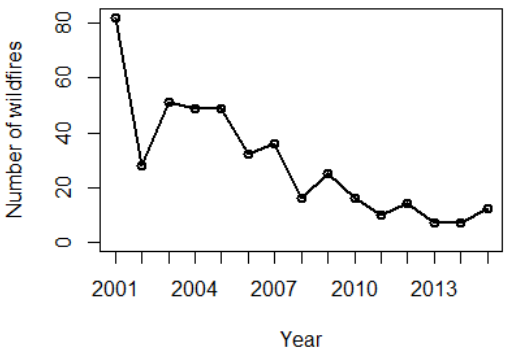}
\caption{\footnotesize{(Left) Forest fire locations in UTM coordinate system (distance in meters), with more than 1 \textit{hectare} of burnt area, recorded during the years $2001$ to $2015$ in the Bouches-du-Rhône county in France. (Right) Number of recorded forest fires per year.}}
\label{pic4}
\end{center}
\end{figure}

Our data set is of the form $(\xi_i, t_i)$,  $i = 1,  \dots, 434$, where $(\xi_i, t_i)$ corresponds to a wildfire with more than $1$ hectare of burnt surface spatially indexed by a DFCI\footnote{district units for fire management strategies, see~\cite{opitz2020}} cell center $\xi_i$ in the Lambert 93 projection system and year $t_i \in \lbrace 2001, \dots, 2015 \rbrace$. To avoid duplicated points we uniformly jittered $\xi_i$ in its DFCI cell. We refer the reader to \cite{gabriel2017} and \cite{opitz2020} for further information on the data.
Whilst forest fires are daily reported, we consider here the yearly scale, as done in many works (see e.g. \cite{serra2012,serra2014a,serra2014b}), because of the small number of reports and to optimize computation time in model fitting and validation steps.
Figure~\ref{pic4} plots locations of forest fires (left panel) and yearly number of occurrences (right panel). It shows some clustering at short and medium spatial distances. Note that there exist two particular areas without any fire occurrences as they correspond to a lake (center) and marshlands (South-West). The number of fires  slightly exponentially decreases in time over the 15 years, mainly due to improvements of fire-fighting resources. 

\begin{figure}[h]
\centering
\includegraphics[width=.45\textwidth]{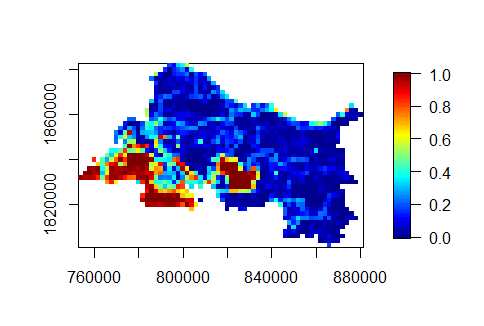}
\includegraphics[width=.45\textwidth]{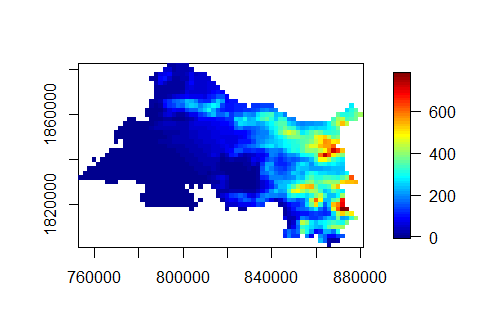}
\includegraphics[width=.45\textwidth]{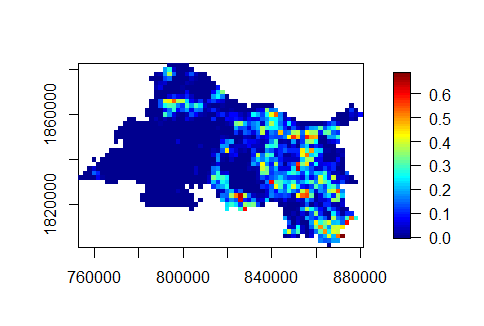}
\includegraphics[width=.45\textwidth]{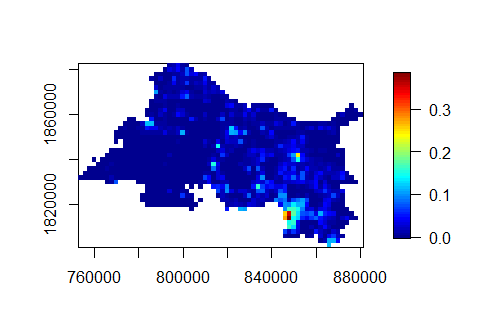}
\includegraphics[width=.45\textwidth]{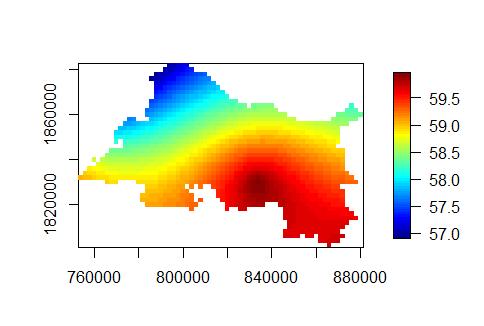}
\includegraphics[width=.45\textwidth]{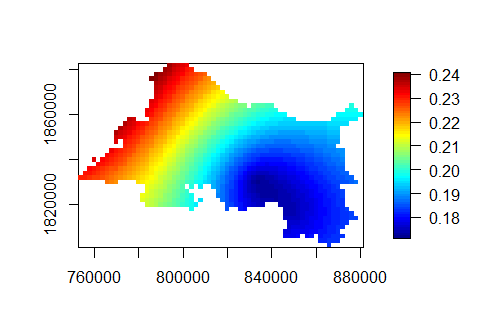}
\caption{\footnotesize{Maps of covariates: water coverage (top left), elevation (top right), coniferous cover (middle left),  building cover (middle right), temperature average (botton left) and square root of precipitation (botton right) in 2001.}}
\label{pic9}
\end{figure}

We consider the same framework as in \cite{gabriel2017} and restrict our attention to the following covariates: water coverage, elevation, coniferous cover and building cover as spatial covariates and temperature average, precipitation as spatio-temporal covariates. Hence, we can consider these covariates as good proxies of the main environmental, climatic and human factors. Maps of covariates are shown in Figure~\ref{pic9} in 2001.

\subsection{Model fitting}

Here we first estimate the spatio-temporal trend and then fit the multi-scale spatio-temporal Geyer model to forest fire occurrences.

\subsubsection{Spatio-temporal trend estimation}

We express the spatio-temporal trend~(\ref{Eq.7}) as $\lambda (\xi,t)=\beta \mu (\xi,t)$ where $\log \mu(\xi,t)$ is assumed to linearly depend on covariates: 
\begin{equation}
\log \mu(\xi,t)=\beta_0+\sum_{k=1}^{4}\beta_k^S Z_k^S(\xi) + \sum_{l=1}^{2} \beta_l^{ST} Z_l^{ST}(\xi,t)+\alpha t
\label{Eq.371}
\end{equation}
with $Z_k^S (\xi)$, $k=1,\dots,4$, the spatial covariates, $Z_l^{ST}(\xi,t)$, $l=1,2$, the spatio-temporal covariates and $\alpha t$ a decreasing trend of fire counts over time.
the coefficient $\beta$ will be estimated simultaneously with the others regular parameters by the logistic likelihood approach.  
Table~\ref{Tab2} reports the coefficients $\beta_0$, $\beta_k^S$, $\beta_l^{ST}$ and $\alpha$ estimated as in~\cite{gabriel2017} and \cite{opitz2020}. The sign indicates if covariates favour (if positive, like coniferous, building and temperature) or prevent (if negative, like water, elevation, precipitation and time) fire occurrences.  
All covariates are significant and results are consistent with previous works.
\begin{table}[h]
\centering
\caption{Estimated coefficients, standard errors and $p$-values based on two-tailed Student's t-tests of significant differences from zero.}
\label{Tab2}
\vspace{.3cm}
\scriptsize
\begin{tabular}{lllll}
\hline
 Covariates & Coefficients              & Estimates & Standard error & $p$-value \\
\hline
Intercept & $\beta_0$ & 262 &26& $<2\times10^{-16\hspace{1mm}***}$    \\
Water & $\beta_1^S$            & -1.88 &0.29 &5.89$\times10^{-11\hspace{1mm}***}$     \\
Elevation & $\beta_2^S$           & -0.001 &0.0004 &$0.0008^{\hspace{1mm}***}$   \\
Coniferous & $\beta_3^S$     & 0.77 &0.36 &$0.031^{\hspace{1mm}*}$   \\
Building & $\beta_4^S$     & 4 &0.89& 8.08$\times10^{-6\hspace{1mm} ***}$   \\
Temperature & $\beta_1^{ST}$     & 0.37 &0.06 &1.13$\times10^{-10 \hspace{1mm} ***}$   \\
Precipitation & $\beta_2^{ST}$     & -11.3 &1.48& $1.75\times10^{-14 \hspace{1mm}***}$   \\
Time & $\alpha$    & -0.14 &0.001& $<2\times10^{-16 \hspace{1mm}***}$   \\
\end{tabular}
\end{table}

\subsubsection{Parameters estimation}

There is no common method for estimating irregular parameters in spatial or spatio-temporal Gibbs point process models. Here we considered several combinations of ad-hoc values within a reasonable range and select the optimal irregular parameters according to the Akaike's Information Criterion (AIC) of the fitted model. 

\cite{baddeley2006} suggest that the spatial interaction radius $r$ of the Geyer saturation point process should be between 0 and the maximum nearest neighbor distance, about 8000 \textit{meters} for our dataset.
For the temporal radius $q$, 
we consider small values to be in accordance with the natural phenomena of forest fire occurrences. 
Finally, for the saturation parameter $s$, we have $n(C_r^q(\xi_i,t_i);\textbf x)\leq s$ for all $(\xi_i,t_i) \in \textbf x$.
 Hence, for any pair $(r,q)$, we set $s=\max_{1\leq i \leq n}n(C_r^q(\xi_i,t_i);\textbf x)$.

According to the former section, we use the logistic likelihood method and \textit{Algorithm 2} to estimate the regular parameters. We simulate dummy points from an inhomogeneous Poisson point process with intensity $\rho(\xi,t)= C \mu(\xi,t)/\nu$ where $C=4$ by a classical rule of thumb in the logistic likelihood approach and $\nu =2000\times2000\times1$ (area of a DFCI cell multiplied by 1 year).

 We fitted the spatio-temporal multi-scale Geyer point process model for a range of ad-hoc values $(r_j , q_j) \in [0, 8000] \times \{1,2,3,4,5\}$,  and  their corresponding  values of $s_j$, $j = 1, \dots,m$, with varying $m$ in $\{1,2,3,4,5\}$.
The minimum AIC is obtained for the combination given in Table~\ref{param}.
Estimated regular parameters show a strong clustering at short distances, randomness at medium and large distances and weak evidence of inhibition.
\begin{table}
\centering
\caption{Parameter estimates for $m=4$.}
\label{param}
\vspace{.3cm}
\scriptsize
\begin{tabular}{rrrrr}
\hline
\multicolumn{5}{c}{Irregular parameters} \\
 \cline{2-5} 
$r$ &500&2000&5000&7500 \\ 
$q$  &1&2&3&4 \\
$s$ &4&7&27&57 \\
\hline
\multicolumn{5}{c}{Regular parameters} \\
\hline
$\hat \beta = 0.66$ & $\hat{\gamma}_1=2.73$, &$\hat{\gamma}_2=0.93$ & $\hat{\gamma}_3=1.07$ & $\hat{\gamma}_4=0.98$ \\
\hline
   \\
\end{tabular}
\end{table}

\subsection{Model validation}

We validate our fitted model from Monte Carlo tests using statistics based on the spatio-temporal inhomogeneous $K$-function~\citep{gabriel2009}. First, we generate $n_{sim} = 99$ simulations from our fitted hybrid Geyer model~\eqref{Eq.7} by \textit{Algorithm 3} with a burn-in period of $70,000$ steps, representing realizations from our null hypothesis. Then, we compute the spatio-temporal inhomogeneous $K$-function
for the observed and simulated point patterns, denoted respectively by $\hat{K}_{obs}^{inh}(h_s,h_t)$ and $\hat{K}_i^{inh}(h_s,h_t), i \in \{1,...,n_{sim}\}$, with an estimated non-separable intensity function obtained by kernel smoothing.  
For each value of the spatio-temporal distance $(h_s,h_t)$, lower ($L$) and upper ($U$) critical envelopes of the summary statistics are computed
\begin{equation}
L(h_s,h_t)=\min_{1\leq i \leq n_{sim}}  \hat{K}_i^{inh}(h_s,h_t), \quad U(h_s,h_t)=\max_{1\leq i \leq n_{sim}}  \hat{K}_i^{inh}(h_s,h_t).
\label{Eq.40}
\end{equation}
In addition to these local envelopes, we compute local and global $p$-values as in \cite{tamayo2014, siino2018a} in order to respectively detect spatio-temporal distances where the departure from the null hypothesis is the most significant and the overall adequacy of our model.
Let $E(h_s,h_t)$ and $V(h_s,h_t)$ denote the mean and variance of $\left \lbrace \hat{K}_1^{inh}(h_s,h_t), \dots, \hat{K}_{n_{sim}}^{inh}(h_s,h_t), \hat{K}_{obs}^{inh}(h_s,h_t) \right \rbrace$. 
We define the local $p$-value for each pair $(h_s,h_t)$ by
\begin{equation}
p(h_s,h_t)=\frac{1+\sum_{i=1}^{n_{sim}}\mathbbm{1}\{T_i(h_s,h_t)>T_{obs}(h_s,h_t)\}}{n_{sim}+1},
\label{Eq.41}
\end{equation}
where $T_i(h_s,h_t)$ (resp. $T_{obs}(h_s,h_t)$) denotes the local statistic $T$ computed from the $i^{th}$ simulation (resp. the data) at $(h_s,h_t)$. The local statistic is  
defined by
\begin{equation}
T(h_s,h_t)= \sqrt{\frac{(\hat{K}^{inh}(h_s,h_t)-E(h_s,h_t))^2}{V(h_s,h_t)}}.
\label{Eq.421}
\end{equation}

The global test combines the information for all spatial and temporal distances.
We define the test statistic 
\begin{equation}
\tilde T=\int_{0}^{h_{t,max}}\int_{0}^{h_{s,max}} T(h_s,h_t)dh_sdh_t, 
\label{Eq.42}
\end{equation}
where $h_{s,max}$ and $h_{t,max}$ are user-specific maximum spatial and temporal distances which are preferable to choose close
to the (expected) range of interaction of the underlying point process. 
\cite{illian2008} recommends to compare the results for several values of $h_{s,max}$ and $h_{t,max}$.
 The $p$-value of the global test is then given by $$p_{global}=\frac{1+\sum_{i=1}^{n_{sim}}\mathbbm{1}\{\tilde T_i> \tilde T_{obs}\}}{n_{sim}+1}.$$ 

\begin{figure}[h]
\begin{center}
\includegraphics[width=1\textwidth]{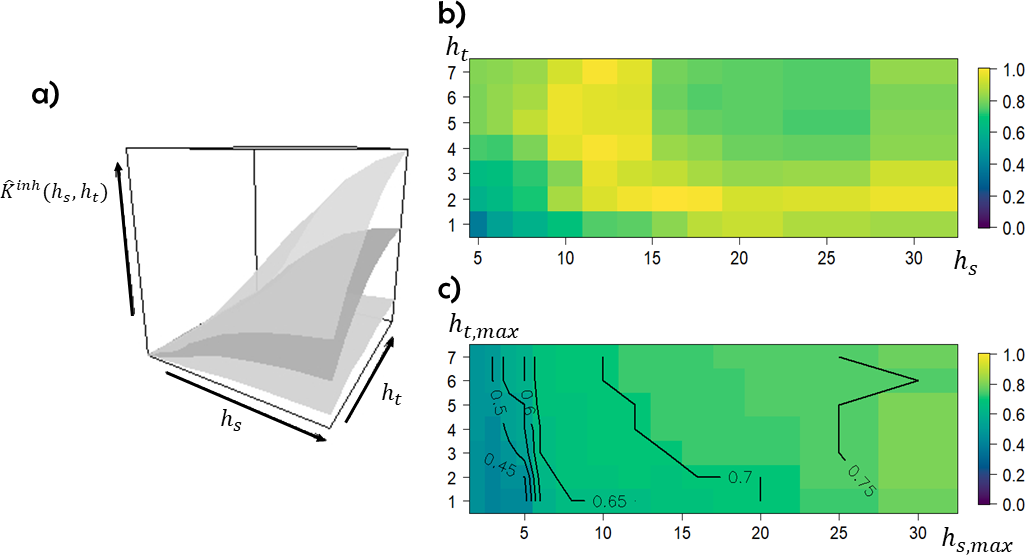}
\caption{\footnotesize{Temporal separations $h_t$ are in \emph{year} and spatial distances $h_s$ are in \emph{kilometer}.
$a)$~Envelopes of the spatio-temporal inhomogeneous $K$-function for the simulated  spatio-temporal  multi-scale Geyer point process  according to the estimated parameters. $b)$~Image plot of the local $p$-value.
$c)$~Image plot of the global $p$-value for any pairs of $(h_{s,max},h_{t,max})$.}}
\label{pic15}
\end{center}
\end{figure}

Figure~\ref{pic15}.$a)$ shows the spatio-temporal inhomogeneous $K$ function computed on our dataset (dark grey) and the envelopes obtained from our hybrid Geyer model (light grey);  $\hat{K}_{obs}^{inh}(h_s,h_t)$ lies inside the envelopes, meaning that the fitted model seems to describe properly the spatio-temporal structure of the data. This is confirmed by local $p$-values at any distances (Figure~\ref{pic15}.$b)$.
Global $p$-values are given in Figure~\ref{pic15}.$c)$ for any combination of  $h_{s,max}$ and $h_{t,max}$. Again, it shows that our fitted model is validated.

Note that we did the same tests for 99 simulations of an inhomogeneous Poisson process with intensity $\mu(\xi,t)/(2000\times 2000 \times 1)$ (\ref{Eq.371}). This model has been rejected at the level $5\%$, with a median global $p$-value equals to 0.04.

\section*{Conclusion}

Due to the capability of Gibbs point processes to cover prevalent structures (inhibition, randomness and clustering), the hybridization approach allows to introduce new Gibbs models combining several structures at different scales. In this paper, we defined the spatio-temporal multi-scale Geyer saturation point process model and detailed the classical statistical inference methods and MCMC simulation techniques that we have extended to the spatio-temporal framework and implemented in {\tt R} code that will be added to the {\tt stpp} package~\cite{gabriel2013}. Our simulation study highlighted a better goodness-of-fit of parameters for the logistic likelihood approach compared to the pseudo-likelihood approach. Finally, we illustrated the interest of using this model on a spatio-temporal dataset of forest fire locations associated with environment covariates. The model validation shows that our model captures the multi-scale interaction structure inherent to forest fire occurrences. 

In this paper, we focused our attention on the definition of a new hybrid Gibbs model, the inference methods and MCMC simulation algorithms that we needed to adapt to the spatio-temporal context. Some of our choices can be discussed and eventually improved in future works, notably in our application to forest fire occurrences which is not presented as an in-depth study but as an illustration of the model fitting on real data.


In our application study, we considered a log-linear form for the trend depending on covariate information. We chose a two-step procedure for estimating, at first, the trend coefficients and then the regular parameters of the interaction function. Our knowledge on forest fire mechanisms guided this choice because the main driver of occurrence locations is the environmental heterogeneity and the secondary one is the interaction phenomena. The trend is estimated at the spatial DFCI scale and at the yearly one, corresponding to our covariate resolution. In that way, we estimated a global trend at a medium scale whereas the interaction parameters are estimated at the point locations and represent a local interaction behavior at a fine scale. 
This procedure could be improved by incorporating variable selection methods, e.g. via regularization~\citep{choiruddin2018,daniel2018}.



Our two-step estimation procedure allows us to provide confidence intervals for the trend coefficients but we have not explored the significance testing of the regular parameters in the interaction function. We notice that some parameters $\gamma_j$ are closed to one and testing departure from one is possible by extending the adjusted composite likelihood ratio test \citep{baddeley2016} to the spatio-temporal framework or by considering parametric bootstrap procedures. Here, we decided to present the testing of the overall goodness-of-fit of our model instead of focusing our work on the pointwise significance of the interaction parameters.


For the choice of irregular parameters, because the likelihood is not differentiable with respect to them, we used a maximum profile likelihood approach based on the logistic likelihood estimation procedure and AIC values for model selection. Introduced for the pseudo-likelihood estimates in \cite{anwar2015} and applied to the logistic likelihood approach by us using the results in \cite{baddeley2014}, this method consists in fixing irregular parameters and maximizing the composite likelihood with respect to the regular ones. This technique is a computationally-intensive method. Thanks to a preliminary spatio-temporal exploratory analysis of the interaction ranges done with the inhomogeneous pair correlation function $g$,  the maximum nearest neighbor distance and the temporal autocorrelation function, we chose few configurations of feasible values for the nuisance parameters $m$, $r_j$, $q_j$ and $s_j$, $j=1\dots m$. Considering more values would be very time-consuming and developing a new estimation method would be a subject in its own right.  


Our model can be used in many fields, like seismology and epidemiology for example, because several mechanisms exhibit interaction between points at multiple scales in space and time. Relying on this work, we can also develop hybrid models with different density structures. Indeed, although it was not necessarily highlighted here, we know that forest fires with large burnt areas avoid future fire occurrences during a vegetation regeneration period. Such cases of strong inhibition may be modeled by hybrid Gibbs point processes with a hardcore component like the hybrid Geyer hardcore point process. We recently extended our work to this model.

\bibliography{mybibfile}

\end{document}